\documentstyle[12pt,a4,fleqn]{article}

\begin{document}

\title{Hypervelocity Impact Fusion with Compressed Deuterium-Tritium 
Targets}

\author{Silviu Olariu \\
Institute of Physics and Nuclear Engineering\\
Department of Fundamental Experimental Physics\\
P.O. Box MG-6, 76900 Magurele, Bucharest, Romania}

\date{}
\maketitle

\abstract
The neutron  yields  observed  in  inertial  confinement  fusion 
experiments for higher convergence ratios are about  two  orders 
of magnitude  smaller  than  the  neutron  yields  predicted  by 
one-dimensional models, the discrepancy being attributed to  the 
development of instabilities. We consider the  possibility  that 
ignition and a moderate gain could  be  achieved  with  existing 
laser facilities if the laser driver energy is used  to  produce 
only  the  radial  compression  of  the  fuel  capsule  to  high 
densities but relatively low temperatures, while the ignition of 
the fusion reactions in the  compressed  fuel  capsule  will  be 
effected   by   a   synchronized   hypervelocity    impact.    A 
positively-charged incident projectile  can  be  accelerated  to 
velocities  of  $3.5\times  10^6$ m s$^{-1}$  ,   resulting   in   
ignition 
temperatures of about 4 keV,  by  a  conventional  low-$\beta$
linac 
having a length of 13 km if deuterium-tritium densities of 570 g 
cm$^{-3}$   could be obtained by laser-driven compression. 
\endabstract

\section{ Introduction }

The standard approach to  inertial  confinement  fusion  is 
based  on  the   deposition   in   the   outer   layers   of   a 
deuterium-tritium fuel capsule of sufficient energy to produce a 
hot central spot of moderate density but high temperature, while 
the remaining fuel mass has a high density but a relatively  low 
temperature \cite{1}. 
Deuterium-tritium fuel densities of 20 to 40  g cm$^{-3}$    have 
been reported in cryogenic target experiments conducted  at  the 
University of  Rochester  \cite{2},  and  compressed  core  densities 
estimated to be 120 g cm$^{-3}$   have been reported in tests conducted 
at the Osaka University \cite{3}.  High  deuterium-tritium  densities 
have  also  been  obtained  at   Lawrence   Livermore   National 
Laboratory and at the Centre d'Etudes de Limeil-Valenton.  There 
is a consensus that ignition and a moderate  gain  for  inertial 
confinement fusion can be achieved with a driver energy  of  the 
order of 1 MJ, subject to stringent uniformity requirements. 

While target gains of the order of  100  are  predicted  by 
one-dimensional simulations  for  a  MJ  laser  input  \cite{4},  the 
neutron  yields  observed  in   inertial   confinement    fusion 
experiments for higher convergence ratios are about  two  orders 
of magnitude smaller than the  neutron  yields  calculated  with 
one-dimensional models,  the  discrepancy  being  attributed  to 
non-uniformities of the drive flux and  to  the  development  of 
instabilities \cite{5}. The  present  goal  of  inertial  confinement 
fusion research is to achieve ignition and a moderate  gain $ G>1 $
\cite{6}. 

In  the  conventional  approach  to  inertial   confinement 
fusion, both the high-density fuel mass and the hot central spot 
are supposed to be produced by  the  deposition  of  the  driver 
energy in the outer layers  of  the  fuel  capsule.  While  high 
densities can be produced indeed, this approach  appears  to  be 
less efficient in igniting the fusion  processes,  as  mentioned 
previously. It is proposed in this paper that the driver  energy 
should be used only to produce the  radial  compression  of  the 
fuel capsule to high densities but relatively low  temperatures, 
while the ignition of fusion reactions in  the  compressed  fuel 
capsules should be  effected  by  a  synchronized  hypervelocity 
impact. The hypervelocity-impact ignition of fusion reactions in 
a compressed fuel capsule is expected to result in higher  gains 
than those actually possible with the conventional approach, for 
the same driver energy. 

In Sec. 2 we describe the general processes involved in the 
hypervelocity-impact ignition of  nuclear  fusion  reactions  in 
laser-compressed fuel pellets. In Sec. 3 we discuss  the  amount 
of  electric  charge  which  can  be  carried  by  an   incident 
projectile and then determine an expression for  the  length  of 
the linac that would accelerate the projectile to  the  required 
velocity.  A  case  study  is  reported  in  Sec.  4,  where  an 
acceleration length of 13 km is obtained on the assumption  that 
deuterium-tritium densities of  570 g cm$^{-3}$   
could be obtained  by 
laser-driven compression. The impact-heating rates are estimated 
in this case to be as high as 1.1 x 10$^{19}$   W cm$^{-2} $ . 
At pressures of 
about 10$^{-11}$    torr in the accelerating column the heating and  the 
ionization  produced  by  the  interaction   of   the   incident 
projectile with the background gas  molecules  is  found  to  be 
within acceptable limits. In Sec. 5 we mention  the  possibility 
to   build   a   hypervelocity   accelerator   and   a    linear 
electron-positron collider as parallel structures  on  the  same 
site, thus sharing the costs  of  the  cryogenic  units,  vacuum 
technique and of the tunnel. An alternative approach is based on 
the generation of an electric-field wave front with the  aid  of 
successively activated switches.

\section{  Hypervelocity-impact  ignition  of   laser-compressed   fuel     
pellets }

We shall consider an incident projectile  consisting  of  a 
spherical shell of radius $r$ and thickness $\delta$, made of a  material 
having a high mechanical tensile strength and a high evaporation 
field. This projectile we suppose to be incident with a velocity 
of 3.5 x 10$^6$  m s$^{-1}$   upon a large-yield fuel target at  rest.  The 
collision of the incident  projectile  and  of  the  large-yield 
target takes place inside a double cylindrical high-Z  case,  as 
shown in Fig. 1. A laser pulse is converted at the walls of  the 
larger  cavity  into  X  rays,  which  compress   the   incident 
projectile and the large-yield target  in  high-density  states. 
The  difference  in  the  compression  times  of  the   incident 
projectile and of the fuel target is  compensated  by  the  fact 
that the X-ray pressure is exerted on  the  incident  projectile 
only for the final part of its path. A second, short laser pulse 
is projected inside the smaller cavity  to  ensure  the  uniform 
illumination of the incident projectile during the entrance into 
the larger cavity. The main laser pulse and the movement of  the 
incident projectile are synchronized  such  that  the  collision 
should take place  when  the  densities  are  the  highest.  The 
collision converts the kinetic energy of the incident projectile 
into thermal energy, the resulting  temperature  being  supposed 
sufficient  to  ignite  the   fusion   reactions   between   the 
deuterium-tritium nuclei contained by the fuel target  at  rest. 
The fusion reactions thus ignited  will  be  propagated  in  the 
entire volume of the large-yield target. 

The possibility  to  use  hypervelocity  impact  to  ignite 
nuclear  fusion  reactions  has  been  previously   investigated 
especially with regard to the type of  the  linear  acceleration 
that would provide kinetic energies of the  order  of  1  MJ  to 
projectiles with a mass of  about  0.1  g  \cite{7}-\cite{11}.  
A  numerical 
one-dimensional   study   of   deuterium-tritium   ignition   in 
impact-fusion targets found necessary shell energies of about 30 
MJ \cite{12}. The magnitude of the large threshold energy  in  impact 
fusion in the absence of laser compression can be  explained  by 
the asymmetry of the pressures exerted during the  collision  of 
two macroscopic bodies. 

Thus, by decoupling the mechanism producing the compression 
of the fuel from the mechanism  producing  the  heating  of  the 
fuel, it  becomes  possible  to  use  the  spherically-symmetric 
pressure exerted by a radiation driver to compress the fuel to a 
high-density state, and  at  the  same  time  to  use  the  high 
energy-transfer rates  characteristic  to  impact  processes  to 
efficiently  heat  the  compressed  fuel,  with  the  aid  of  a 
comparatively small amount of kinetic energy. 

We shall  assume  that  the  velocities  required  for  the 
ignition  of  nuclear   fusion   reactions   are   obtained   by 
accelerating the positively-charged projectiles along  a  series 
of coaxial tubular electrodes connected alternatively to two bus 
bars and  supplied  by  a  radio-frequency  power  source.  This 
arrangement, due to Wideroe \cite{13}, is used when the ratio of  the 
velocity of the charged projectile to the velocity of  light  is 
small compared to 1. The acceleration of macroparticles based on 
an electrostatic charge  carried  by  the  projectile  has  been 
previously regarded as uncompetitive because of  the  length  of 
the accelerator \cite{8},\cite{11},\cite{14}, 
but those estimates have  been  based 
on the presumption of a MJ kinetic energy for a 0.1 g projectile 
incident with a velocity of about $1.5 \times 10^5$  m s$^{-1}$  .  
However,  if 
the fuel target at rest is  driven  in  the  high-density  state 
before  the  collision,  the   kinetic   energy   threshold   is 
substantially lowered. Moreover,  since  the  required  incident 
velocity of about $3.5 \times 10^6$  m s$^{-1}$,  
which  would  result  in  an 
ignition temperature of about 4 keV, is one order  of  magnitude 
larger  than  the   velocity   used   in   the   afore-mentioned 
evaluations, then the mass of the  incident  projectile  becomes 
very small, of the order of 10$^{-6}$  g,  and  for  masses  of  this 
magnitude the efficient acceleration approach is that  based  on 
the electrostatic charge. 

\section{ Acceleration of the incident projectile }

We assume that the incident projectile is a  hollow  sphere 
of radius $r$ and thickness $\delta$, of aspect ratio 
\begin{equation}
A=r/\delta,
\label{1}			
\end{equation} 
This moving hollow sphere is compressed inside the case shown in 
Fig. 1 to a full sphere of radius $r_0$ , so that  the  material  of 
the projectile undergoes a compression by a factor 
\begin{equation}
P=3r^2\delta/r_0^3,
\label{2}
\end{equation}
At the moment of the  impact,  the  density  of  the  compressed 
deuterium-tritium fuel of the target at rest  is  $n_0$ ,  the  fuel 
being initially compressed by a factor 
\begin{equation}
F=n_0/n_s,
\label{3}
\end{equation}
where $n_s  = 4.5 \times 10^{22}$   cm$^{-3}$    
is  the regular solid-state density 
of equimolar deuterium-tritium. We  shall  express  the  various 
quantities  in  this  paper  in  terms  of   the   dimensionless 
parameters $A, P, F.$

We  assume  moreover  that  the  kinetic  energy   of   the 
compressed projectile of radius $r_0$  is converted with  a  certain 
efficiency $\eta_{tot}$ into thermal energy at  temperature  $T_0$   of  the 
compressed deuterium-tritium fuel over  a  sphere  of  the  same 
radius $r_0$ , situated in the impact region. According to Brueckner 
and Jorna \cite{15}, ignition  of  the  compressed  deuterium-tritium 
fuel will take place if 
\begin{equation}
r_0=r_c/F ,
\label{4}
\end{equation}
where $r_c  = 2.81 \times 10^{-2}$ m  when $ kT_0$ =  4  keV,  
and  the  initial 
thermal energy $E_{th}$ to produce ignition of the  deuterium-tritium 
sphere of radius $r_0$  is ,
\begin{equation}
E_{th}=E_u/F^2 ,
\label{5}
\end{equation}
where $E_u= 7.99 \times 10^9$  J for the same initial temperature of  $kT_0$  
= 4 keV. Then  from  Eqs.  (1)-(5),  the  initial  radius  of  the 
projectile is  
\begin{equation}
r=\frac{r_c}{F}\left(\frac{AP}{3}\right)^{1/3} ,
\label{6}
\end{equation}
and the thickness of the spherical shell is                     
\begin{equation}
\delta=\frac{r_c}{F}\left(\frac{P}{3A^2}\right)^{1/3} .
\label{7}
\end{equation}
We shall now determine the amount of electric   charge   $q_\delta$   
which  can be carried by the incident projectile. The  variables 
which limit the amount of positive electric charge which can be 
carried by a hollow sphere are the  mechanical  strength,  which 
opposes the electrostatic  repulsion  forces,  and  the  surface 
electric field, which determines the field-evaporation  and  the 
field-ionization phenomena. It can be  shown  that  the  maximum 
electric field $E_\delta$  which  can  be  mechanically  sustained  by  a 
spherical shell made of a material of  tensile  strength  $\sigma$,  of 
radius $r$ and thickness $\delta\ll r$, is given by 
\begin{equation}
E_\delta=(4\delta\sigma/\epsilon_0 r)^{1/2} ,
\label{8}
\end{equation}
in SI units. The ideal tensile strength of  a  material  can  be 
estimated as \cite{16} 
\begin{equation}
\sigma=E/10 ,
\label{9}
\end{equation}
where $E$ is the Young's modulus. At the same time,  the  electric 
field at the surface of  the  projectile  must  not  exceed  the 
field-evaporation  limit.  Large  surface  electric  fields  are 
encountered in field-ion microscopy \cite{17},\cite{18}, where  observations 
are   frequently   made   at   fields    comparable    to    the 
field-evaporation limit, of the order of 
$5 \times  10^{10} $   V  m$^{-1}$  .  
For spherical shells, the more stringent condition is that  ensuring 
the mechanical stability of the charged shell, Eq. (8). Then the 
maximum amount of positive electric charge which can be  carried 
by the fuel capsule is 
\begin{equation}
q_\delta=4\pi\epsilon_0 r^2 E_\delta ,
\label{10}
\end{equation}
which can be written with the aid of Eqs. (6)-(8) as 
\begin{equation}
q_\delta=\frac{8\pi}{3^{2/3}}\epsilon_0^{1/2}\sigma^{1/2}A^{1/6}P^{2/3}
\frac{r_c^2}{F^2} .
\label{11}
\end{equation}

We shall consider that the total efficiency $\eta_{tot}$ with which 
the kinetic energy of the  incident  projectile  is  transformed 
into thermal energy of the compressed deuterium-tritium fuel  is 
a product of three factors. The first of  these  factors,  which 
can be obtained from the conservation of the energy and momentum 
during the impact, is $(1+m_\delta/m_0)^{-1}$  , where 
\begin{equation}
m_\delta=4\pi r^2\delta \rho_Z 
\label{12}
\end{equation}
is the mass of the incident projectile of density $\rho_Z$, and 
\begin{equation}
m_0=\frac{4\pi r_0^3}{3}m_{DT}n_0 
\label{13}
\end{equation}
is the initial mass of deuterium-tritium fuel whose  temperature 
is raised to $kT_0$  as a result of the impact, 
$m_{DT}   = 4.17  \times 10^{-27}$  kg  
being the average of the masses of a deuterium nucleus and of 
a tritium nucleus. The second factor takes into account the fact 
that   the   thermal   energy   is   distributed   between   the 
deuterium-tritium nuclei, the nuclei  of  the  material  of  the 
projectile and the corresponding electrons, and is given by 
\begin{eqnarray}
\left(1+\frac{3}{2}(Z+1)\frac{r^2\delta}{r_0^3}\frac{n_Z}{n_0}\right)^{-1}
\nonumber
\end{eqnarray}
where the number $Z$ characterizes the material of the projectile. 
The third factor is an empirical efficiency $\eta$, by which we  take 
into account the deviations from the assumptions of  our  model. 
Then the total efficiency is of the form 
\begin{equation}
\eta_{tot}=\eta\left(1+\frac{P}{F}\frac{\rho_Z}{\rho_s}\right)^{-1}
\left(1+\frac{1}{2}(Z+1)\frac{P}{F}\frac{n_Z}{n_s}\right)^{-1} ,
\label{14}
\end{equation}
where $\rho_s=m_{DT}n_s$  is the density of solid deuterium-tritium, 
$\rho_s$= 0.19 g cm$^{-3}$  . 

The  potential  difference $\Phi_\delta$ accelerating  the  incident 
projectile of charge $q_\delta$   to  an  energy  which,  converted  with 
efficiency  $\eta_{tot}$,  results  in  a  thermal  energy  $E_{th}$ at a 
temperature $kT_0 $ = 4 keV of the compressed deuterium-tritium fuel 
is 
\begin{equation}
\Phi_\delta=\frac{E_{th}}{\eta_{tot}q_\delta} ,
\label{15}
\end{equation}
and can be expressed in terms of the dimensionless parameters 
$A, P, F $ as 
\begin{equation}
\Phi_\delta=\frac{3^{2/3}}{8\pi\eta}
\left(1+\frac{P}{F}\frac{\rho_Z}{\rho_s}\right)
\left(1+\frac{1}{2}(Z+1)\frac{P}{F}\frac{n_Z}{n_s}\right)
\frac{E_uA^{-1/6}P^{-2/3}}{\epsilon_0^{1/2}\sigma^{1/2}r_c^2} ,
\label{16}
\end{equation}

As mentioned previously, we suppose that the positively-charged 
projectiles are accelerated to the  required  energy  by 
the electric field extant  in  the  gaps  between  a  series  of 
tubular electrodes connected to a radio-frequency source.  Since 
the low-$\beta$ values result in relatively low  applied  frequencies, 
the  electric  field  extant  in  a  gap  of  length  $d$  between 
successive electrodes can be estimated as the  static  breakdown 
field for the same electrodes \cite{19}, 
\begin{equation}
E=Kd^{-1/2}.
\label{17}
\end{equation}

The average accelerating field $E_{av}$   is reduced by a factor 
$2^{3/2}/\pi$ with  respect  to  $E$  because  of  the  sinusoidal  time 
dependence, and moreover it is reduced by a factor of  1/2  when 
we take into account the length of  the  drift  tubes,  supposed 
equal to the gap length, 
\begin{equation}
E_{av}=(2^{1/2}/\pi)Kd^{-1/2} .
\label{18}
\end{equation}
where the value $K = 2.4 \times 10^6$,  in  SI  units,  corresponds  to 
copper electrodes with an area of a few square  centimeters.  We 
shall determine the radius $a$ of the tubular electrodes from  the 
requirement that the electric field of the charged projectile at 
the distance $a$ from its center should not exceed  the  breakdown 
field, Eq. (17), for the distance $a$, 
\begin{equation}
a=(q_\delta/4\pi\epsilon_0 K)^{2/3} .
\label{19}
\end{equation}
and shall consider that the gap spacing $ d$ is a certain  multiple 
$N$ of the diameter $2a$ of the tube, 
\begin{equation}
d=2Na
\label{20}
\end{equation}

The average electric field can then be written in terms  of 
$A, P, F$ as 
\begin{equation}
E_{av}=\frac{3^{2/9}K^{4/3}\epsilon_0^{1/6}F^{2/3}}
{2^{1/3}\pi N^{1/2}\sigma^{1/6}r_c^{2/3}A^{1/18}P^{2/9}} ,
\label{21}
\end{equation}
so that the length of the accelerator $L_\delta  = \Phi_\delta/E_{av}$   is
\begin{equation}
L_\delta=\frac{3^{4/9}N^{1/2}}{2^{8/3}\eta}
\left(1+\frac{P}{F}\frac{\rho_Z}{\rho_s}\right)
\left(1+\frac{1}{2}(Z+1)\frac{P}{F}\frac{n_Z}{n_s}\right)
\frac{E_u
A^{-1/9}P^{-4/9}F^{-2/3}}{\epsilon_0^{2/3}\sigma^{1/3}r_c^{4/3}K^{4/3}} ,
\label{22}
\end{equation}

The radius of the drift tubes is 
\begin{equation}
a=\frac{2^{2/3}}{3^{4/9}}
\frac{\sigma^{1/3}r_c^{4/3}A^{1/9}P^{4/9}}{\epsilon_0^{1/3}K^{2/3}F^{4/3}} ,
\label{23}
\end{equation}
and the electric field of the projectile at distance $a$ is 
\begin{equation}
E(a)=\frac{3^{2/9}}{2^{5/6}N^{1/2}}\frac{\epsilon_0^{1/6}K^{4/3}F^{2/3}}
{\sigma^{1/6}r_c^{2/3}A^{1/18}P^{2/9}}
\label{24}
\end{equation}
The electric field at the surface of the incident projectile, 
$E_s=q_\delta/4\pi\epsilon_0r^2$, is
\begin{equation}
E_s=2\epsilon_0^{-1/2}\sigma^{1/2}A^{-1/2} .
\label{25}
\end{equation}

The mass of the incident projectile, previously written  in 
Eq. (12), is 
\begin{equation}
m_\delta=\frac{4\pi}{3}\frac{r_c^3}{F^3}P\rho_Z .
\label{26}
\end{equation}
The final energy of the incident projectile is 
\begin{equation}
E_\delta=E_{th}/\eta_{tot} ,
\label{27}
\end{equation}
and the final velocity $v$ of the projectile, calculated from  the 
relation $m_\delta  v^2/2=E_\delta$, is 
\begin{equation}
v=\frac{3^{1/2}}{2^{1/2}\pi^{1/2}\eta^{1/2}}
\left(1+\frac{P}{F}\frac{\rho_Z}{\rho_s}\right)^{1/2}
\left(1+\frac{1}{2}(Z+1)\frac{P}{F}\frac{n_Z}{n_s}\right)^{1/2}
\frac{E_u^{1/2}F^{1/2}P^{-1/2}}{\rho_Z^{1/2}r_c^{3/2}\eta^{1/2}} .
\label{28}
\end{equation}

The average force acting on the projectile, calculated as  
$F_\delta=q_\delta E_{av}$ , is
\begin{equation}
F_\delta=\frac{2^{8/3}}{3^{4/9}N^{1/2}}\epsilon_0^{2/3}\sigma^{1/3}r_c^{4/3}
K^{4/3}\frac{A^{1/9}P^{4/9}}{F^{4/3}} ,
\label{29}
\end{equation}
and the average acceleration of the projectile is 
\begin{equation}
a_\delta=\frac{2^{2/3}3^{5/9}}{\pi N^{1/2}}\frac{\epsilon_0^{2/3}
\sigma^{1/3}K^{4/3}A^{1/9}F^{5/3}}{r_c^{5/3}P^{5/9}\rho_Z}
\label{30}
\end{equation}
The time of transit of the projectile through  the  accelerator, 
calculated as $t_\delta  = (2L_\delta /a_\delta )^{1/2}$    , is 
\begin{equation}
t_\delta=\frac{\pi^{1/2}N^{1/2}}{2^{7/6}3^{1/18}\eta^{1/2}}
\left(1+\frac{P}{F}\frac{\rho_Z}{\rho_s}\right)^{1/2}
\left(1+\frac{1}{2}(Z+1)\frac{P}{F}\frac{n_Z}{n_s}\right)^{1/2}
\frac{E_u^{1/2}r_c^{1/6}P^{1/18}\rho_Z^{1/2}}
{\epsilon_0^{2/3}\sigma^{1/3}K^{4/3}A^{1/9}F^{7/6}} ,
\label{31}
\end{equation}
Finally the specific charge of the projectile is 
\begin{equation}
\frac{q_\delta}{m_\delta}=
\frac{24^{1/3}\epsilon_0^{1/2}\sigma^{1/2}A^{1/6}F}{r_cP^{1/3}\rho_Z} .
\label{32}
\end{equation}

\section{ Case study }

If we examine the expressions of  the  acceleration  length 
$L_\delta$ , Eq. (22), and of the potential difference $\Phi_\delta$, 
Eq.  (16),  we 
see that small values of the density $\rho_Z$ and large values of  the 
tensile strength $\sigma$ of the material of the projectile  result  in 
low values of $L_\delta$  and $\Phi_\delta$. 
We shall consider that the material  of 
the projectile is beryllium, with $Z=4, \rho_Z= 1.85$ g cm$^{-3}$    and  an 
ideal tensile strength, Eq. (9), $\sigma= 2.96\times 10^{10} $   
N  m$^{-2}$.  
The values of the  acceleration  length  $L_\delta$   and  of  the  potential 
difference $\Phi_\delta$ have been represented in Fig. 2  and  respectively 
Fig. 3 as functions of the ratio $P/F$,  for $ F$=1000  and $ F$=3000, 
taking the aspect ratio, Eq. (1), $A$=10, the  empiric  efficiency 
introduced in Eq. (14) $\eta$=0.5 and the parameter $N$ defined in  Eq. 
(20) $N$=2. The length $L_\delta$  is minimized in the case of a  beryllium 
projectile when $P/F$ = 0.0348, and the potential $\Phi_\delta$ is minimized
when $P/F$ = 0.0611, while $F$ is as large as possible.

We  shall discuss two cases, one when the parameters $P, F$
defined in Eqs. (2),(3) are $P=20, F=1000,$ and a second  case  when 
$P=60, F=3000$, the other parameters being $A=10,  \eta=0.5,  N=2$.  We 
kept the compression factor $P$ of  the  material  of  the  moving 
projectile as small as possible, so that $P/F$ =  0.02,  resulting 
in values of $L_\delta$  and $\Phi_\delta$ 
slightly greater than the minimum values. 
We  assume  that  the  compressed  deuterium-tritium   fuel   is 
impact-heated to a temperature $kT_0$  = 4 keV, so that $r_c  = 2.81 \times
10^{-2}$ m, 
$E_u= 7.99 \times 10^9$   J,  as  mentioned  previously  in  Eqs. 
(4),(5). 

In the case when  $P = 20$ and $F$ = 1000, the radius of the 
projectile is $r$ = 114 $\mu$m and  the  thickness  of  the  spherical 
shell is $\delta$ = 11 $\mu$m. The density of the beryllium  projectile  is 
increased before the impact by a factor $P$ = 20, to  37  g  cm$^{-3}  . $
The radius of the compressed fuel region, of initial density  $n_0$  
= 190 g cm$^{-3}$  , and which is heated by impact to a temperature $kT_0$  
= 4 keV, is $r_0$  = 28 $\mu$m, and the initial thermal  energy  of  the 
deuterium-tritium fuel is $E_{th}$= 7990 J. The  kinetic  energy  of 
the  projectile  is  converted  into  thermal  energy   of   the 
compressed deuterium-tritium fuel with a total efficiency $\eta_{tot}$= 
0.36. The charge carried by the projectile is $q_\delta  = 5.28  
\times  10^{-8}$   
C, the accelerating potential is $\Phi_\delta$= 410 GeV, and  the  average 
accelerating field is $E_{av}   = 9.27 \times 10^6$  V m$^{-1}$  . 
The electric field 
at the surface of the beryllium projectile is $E_s  = 3.65 \times 10^{10}$
V m$^{-1}$  , compared to the desorption field of beryllium of  5.36  $\times 
10^{10}$   V m$^{-1}$   \cite{20}. The mass of the incident  projectile  is  
$m_\delta = 3.43 \times 10^{-9}$   kg, 
the final velocity of the projectile is $v = 3.55 
\times 10^6$ m s$^{-1}$  , and the final energy of the projectile is 
$E_\delta=  2.16 
\times 10^4$  J. The radius of a drift tube is $a$=3.39 mm,  the  distance 
between the tubes is $d$ = 1.35 cm, and the electric field of  the 
projectile at distance $a$ is $E(a) = 2.06 \times 10^7$  V m$^{-1}$  . 
The  length $L$
of the accelerator is $L_\delta$  = 44.3 km. The  average  force  on  the 
projectile is $F_\delta$  = 0.49  N,  the  average  acceleration  of  the 
projectile is $a_\delta  = 1.42 x 10^8$  m s$^{-2}$, 
and the time of transit  of 
the projectile through the  accelerator  is  $t_\delta$   =  25  ms.  The 
specific charge of the projectile is $q_\delta /m_\delta$  = 15.3  
C  kg$^{-1}$  .  
The frequency of the accelerating field, calculated  as  $f  =  v/4d$, 
increases from 184 kHz at the low-velocity end, for  an  initial 
velocity of the projectile of $10^4$   m  s$^{-1}$  ,  to  65  MHz  at  the 
high-velocity end. For 90  \%  of  the  accelerator  length,  the 
frequency will be, however, greater than 20 MHz.

In the case when  $P$ = 60, $ F  =  3000$,  the  radius  of  the 
projectile is $r$=54 $\mu$m and the thickness of the spherical shell  is 
$\delta$= 5.4 $\mu$m. The density of the beryllium projectile is increased 
before impact by a factor $P$ = 60, to 111 g cm$^{-3}$  . The  radius  of 
the compressed fuel region, of initial density $n_0  = 570$ g  cm$^{-3}$  , 
and which is heated by impact to a temperature $kT_0$  = 4  keV,  is 
$r_0$   =  9.3  $\mu$m,  and  the  initial   thermal   energy   of   the 
deuterium-tritium fuel is $E_{th}$= 887 J. The kinetic energy of the 
projectile is converted into thermal energy  of  the  compressed 
deuterium-tritium fuel with a total efficiency $\eta_{tot}$= 0.36,  the 
same as in the previous case because $\eta_{tot}$ depends  only  on  the 
ratio $P/F$. The charge carried by the projectile is 
$q_\delta  =  1.22  \times 
10^{-8}$   C, the accelerating potential is $\Phi_\delta$=  197  
GeV,  and  the 
average accelerating field is $E_{av}    =  1.51  \times  10^7$   
V m$^{-1}$  .  The 
electric field at the surface of the beryllium projectile is  $E_s  
= 3.65 \times 10^{10}$   V m$^{-1}$  , the same as in the previous case.  
The mass 
of the incident projectile is $m_\delta  = 3.82 \times 10^{-10}$ kg,  
the  final 
velocity of the projectile is $v = 3.55 \times 10^6$  m s$^{-1}$  , 
the same  as 
previously, and the final energy of the projectile is $E_\delta$=  2409 
J. The radius of a drift tube is  $a$  =  1.28  mm,  the  distance 
between the tubes is $d$ = 5.1 mm, and the electric field  of  the 
projectile at distance $a$ is $E(a) = 3.35 \times 10^7$  V m$^{-1}  . $
The  length 
of the accelerator is $L_\delta$  = 13.0 km. The  average  force  on  the 
projectile is $F_\delta$  = 0.18  N,  the  average  acceleration  of  the 
projectile is $a_\delta$  = $4.82 \times 10^8$  m s$^{-2}$  , 
and the time of transit  of 
the projectile through the accelerator  is  $t_\delta$   =  7.3  ms.  The 
specific charge of the projectile is $q_\delta /m_\delta$  = 31.9  
C~kg$^{-1}$  .  
The frequency of the accelerating field, calculated  as  $f  =  v/4d$, 
increases from 489 kHz at the low-velocity end, for  an  initial 
velocity of the projectile of $10^4$  m  s$^{-1}$  ,  to  173  MHz  at  the 
high-velocity end. For  90  \%  of  the  accelerator  length  the 
frequency will be, however, greater than 54 MHz. 

In order that the collision of the incident projectile  and 
of the fuel target should take place  when  the  densities  have 
their highest values it is necessary that 
\begin{equation}
v \tau = D - r_0 - R_c ,
\label{33}
\end{equation}
where $\tau $ is the time required for the compression of the incident 
projectile, $D$ is the distance from the projectile entrance  into 
the large cavity to the center of the target at rest,  as  shown 
in Fig. 1,  $r_0, R_c$  are respectively the compressed radii of the 
incident projectile and of the target at  rest,  and  $v$  is  the 
velocity of the projectile given by Eq.  (28).  Since  the  time 
required for the compression of the relatively small capsule  is 
of the order of $\tau$= 1 ns, the product $v\tau$ is of the order of $v\tau$= 
3.5 mm. If the lifetime  $\tau_c$ of  the  compressed  state  of  the 
projectile is of the order of $\tau_c$= 0.1 ns, the position  of  the 
projectile before collision must be defined with an accuracy  of 
less than about $v\tau_c$ =  0.35  mm.  The  time  required  for  the 
compression of the large-yield target can be expected  to  be  o 
the order of 10 ns, which means that the main  driver  pulse  is 
applied when the distance between the projectile and the  target 
is of about 3.5 cm. The duration of the secondary  laser  pulse, 
applied in the small cavity when the projectile enters into  the 
large cavity, can be estimated as $4r/v$, which is 128 ps for  the 
114 $\mu$m projectile and 61 ps for the 54 $\mu$m  projectile.  This  is 
only  an  order-of-magnitude  estimate,  and  further   detailed 
analysis is required to determine  whether  the  moment  of  the 
impact can actually be synchronized with the  moments  when  the 
incident projectile and the fuel target are in the  high-density 
states. 

The  impact-heating  of  the  fuel,  produced  during   the 
collision  of  the  compressed  incident  projectile  with   the 
compressed target at rest, takes place in a time estimated as 
$2 r_0 /v_0$  , where $v_0  = (3kT_0 /m_{DT})^{1/2}$    
is the  thermal  velocity  of 
the fuel nuclei at the temperature $kT_0$  = 4 keV, 
$v_0  = 6.78 \times  10^5$  
m s$^{-1}$  . The impact-heating time of the fuel sphere of  radius  $r$  
is then 82.8 ps for the 114 $\mu$m projectile, and 27.6 ps  for  the 
54 $\mu$m projectile.  In  order  to  estimate  the  energy-transfer 
rates, we can consider  that  the  conversion  of  the  incident 
kinetic energy into heat is  effected over  the  cross-sectional 
area $\pi r_0^2$   of  the  compressed  incident  projectile.  Then  the 
impact-heating rate for  the  114  $\mu$m  projectile,  obtained  by 
dividing  the  7990  J  of  thermal  energy  to  the   82.8   ps 
impact-heating time and the area $2.48 \times 10^{-5}$   cm$^2$ , 
would be  $3.88\times 10^{18}$ 
W cm$^{-2}$  ,  and  the  impact-heating  rate  for  the  54 $\mu$m 
capsule, obtained by dividing the 887 J of thermal energy to the 
27.6 ps impact-heating time and the cross-sectional area $2.75 \times 
10^{-6}$   cm$^2$ , would be $1.16 \times 10^{19}$  W cm$^{-2}$  . 
The very high values of the 
energy-transfer rates show  that  the  hypervelocity  impact  is 
appropriate indeed for the ignition of the fusion processes, but 
we emphasize that these impact-heating rates are  possible  only 
in conjunction with the laser precompression of the fuel. 

The fusion processes ignited in the region of  impact  will 
be propagated in the entire volume of the large-yield target  at 
rest. Since  the  burning  front  advances  supersonically,  the 
hydrodynamic expansion associated with the eccentric position of 
the ignition region appears to be of secondary  importance.  The 
hypervelocity-impact ignition of large-yield fuel  targets  does 
not require additional accelerating length, because the ignition 
criterion, Eq. (5), involves the compression $F = n_0 /n_s$ , and does 
not depend on the total energy yield. 

In the system of reference of the projectile, there  is  an 
oscillatory  electric  field  which  produces   an   alternating 
electric polarization of the projectile. In  order  to  estimate 
the heating produced by the electric  currents  associated  with 
the motion of electricity  in  the  beryllium  shell,  we  shall 
regard the metallic shell as a series $RC$ circuit, the resistance 
$R$ being of the order of $R = \rho_{Be}/6\delta$ and the  
capacitance  of  the 
order of $C =\pi\epsilon_0 r$, where $\rho_{Be}$ 
is the  resistivity  of  beryllium. 
The amplitude of the oscillatory potential difference is of  the 
order of $E_{av}  r$ and the frequency of the order of $\omega=\pi v/d$.  
Then the power dissipated in the beryllium shell is of the order of 
$\pi^4\epsilon_0^2\rho_{Be}r^4 v^2 E_{av}^2/12 \delta d^2$.
For   a   room-temperature    electrical 
resistivity  of  beryllium $\rho_{Be}=4.0\times  10^{-8} \Omega$ m,  and   a 
room-temperature specific heat of beryllium of 1.8  J  g$^{-1}$ K$^{-1}$ , 
the temperature increase  of  the  beryllium  shell  during  the 
acceleration will be of about $8.9 \times  10^{-6}$ K  for  the  114 $\mu$m 
projectile and $4.9 \times 10^{-5}$ K for the 54 $\mu$m projectile. The  small 
temperature increase is due  to  the  small  dimensions  of  the 
projectile. 

A more important source of heating is the friction with the 
molecules of the low-pressure gas  extant  in  the  acceleration 
column. If we consider that  each  gas  molecule  contributes  a 
thermal energy $mv^2 /2$  by  colliding  with  the  projectile,  the 
energy transferred to the beryllium shell over  an  acceleration 
length $L_\delta$  will be $(mv^2/2)\pi r^2 L_\delta n$, 
where $m$ is the mass  of  a  gas 
molecule and $n$ is the concentration of these  molecules.  If  we 
assume that the background  air  pressure  is  $10^{-11}$  torr,  the 
temperature increase of the beryllium shell will be 31 K for the 
114 $\mu$m projectile, and 19 K for the 54 $\mu$m projectile. 

We can estimate the charge collected by the projectile  via 
the collisions with the background air molecules as 
$eZ_{air}\pi r^2 L_\delta n$. 
For a residual pressure of $10^{-11}$    torr, and for $Z_{air} = 14.4$,  the 
charge collected by the 114 $\mu$m  projectile  is  $1.47 \times  10^{-9}$ C 
compared to the total charge $q_\delta  = 5.28 \times 10^{-8}$   C, 
and the  charge 
collected by the 54 $\mu$m projectile is $1.00 \times 10^{-10}$  C 
compared  to  the total charge $q_\delta  = 1.22 \times 10^{-8}$   C. 
                                       
We mention that the electrostatic energy $q_\delta^2 /4\pi\epsilon_0 r_0$   
of  the projectile of charge $q_\delta$ , 
compressed to a sphere of radius $r_0$ , is 
0.89 J for the 114 $\mu$m projectile, and  0.14  J  for  the  54  $\mu$m 
projectile. Thus, the fact that the projectile is  charged  does 
not significantly alter the process of compression from radius $r$ 
to radius $r_0$ .

\section{ Accelerator choices }

The radio-frequency power required to activate a 44 km or a 
13 km  accelerator  can  be  reduced  to  acceptable  values  by 
operating  at  superconducting   temperatures.   Although   very 
different in scope and realization, we  mention  at  this  point 
that several projects of linear electron-positron  colliders  at 
energies  in  the   500   GeV   region   are   presently   under 
consideration, the active length being 6.6  km  for  the  CERN's 
Linear Collider CLIC, and 20  km  for  the  TeV  Superconducting 
Linear Accelerator TESLA \cite{21}. If extended to 2 TeV, the  length 
of the CLIC accelerator would be 2 $\times$ 16 km \cite{22}. One possibility 
would be to build  a  hypervelocity  accelerator  and  a  linear 
electron-positron collider as parallel structures  on  the  same 
site, thus sharing the costs  of  the  cryogenic  units,  vacuum 
technique  and  of  the   tunnel.   The   case   of   a   linear 
electron-positron collider would be strengthened if parts of the 
hardware could be used for an applied research project like  the 
hypervelocity-impact ignition of fusion processes. 

If  we  assume  that  10  incident  projectiles  would   be 
accelerated per second for energy production,  the  accelerating 
structure  is  amenable  to  pulsed  and   modular   excitation, 
improving the energy efficiency of the accelerator,  so  that  a 
charged projectile would actually  be  under  the  action  of  a 
radio-frequency wave front. 

The fact that a single macroscopic particle is  accelerated 
at a time in a hypervelocity  accelerator  renders  possible  an 
alternative design which avoids the use of radio-frequency power 
altogether. In this design, the accelerating line consists of  a 
series of coaxial tubular electrodes and one bus bar held  at  a 
static high positive potential, of  the  order  of $ 10^5$   V,  the 
accelerated projectile moving  along  the  common  axis  of  the 
tubes. Each drift tube is connected to the high-voltage bar  via 
a gas-insulated, three-electrode spark gap having the  structure 
shown  schematically  in  Fig.  4.  The  upper  electrode  T  is 
connected to the drift tube, the lower electrode A is  connected 
to the high-voltage bar, and the electrode F  is  lying  in  the 
proximity of T and has initially the same potential  as  T,  for 
example a zero potential. When the positively-charged projectile 
is inside the drift tube,  the  electrode  T  becomes  polarized 
positively with respect to F, and that initiates a spark  across 
the close gap between T and F. The ionization thus created  then 
initiates the main discharge between T and  A,  at  the  end  of 
which the potential of the electrode T and of the drift tube  to 
which it is connected  becomes  highly  positive.  The  electric 
field between the drift tube thus activated and  the  next  tube 
accelerates the projectile toward the latter tube, which is then 
activated when  the  projectile  passes  through  it,  and  this 
process of positive charging of the drift tubes is continued all 
along the path of the projectile. After the acceleration of  the 
projectile is completed, the drift  tubes  are  discharged,  the 
potential of the bus bar restored to its initial  high  positive 
value, and another projectile can then be accelerated along  the 
line. 

We mention that a 10  m  long  linac  of  this  type  would 
accelerate the 114 $\mu$m projectile  considered  in  Sec.  4  to  a 
velocity of 53 km s$^{-1}$  , and the 54 $\mu$m projectile to a velocity of 
98 km s$^{-1}$  .

Another possibility is to excite the accelerating structure 
by highvoltage pulses of a few picosecond duration, produced by 
photodiode switches triggered from laser pulses \cite{23}. 

The costs of current linac projects are estimated at  about 
1 M\$ per GeV \cite{24}, and it is possible that less  stringent  beam 
requirements should significantly reduce these costs. While  the 
single-particle linear accelerator envisaged  in  this  work  is 
very  different  from  the   afore-mentioned   colliders,   this 
comparison  suggests  however  that  the  cost  of  a  500   GeV 
hypervelocity  accelerator  for  impact-fusion  research   might 
represent  a  relatively  small  fraction  of  the  cost  of   a 
reference   project    like the   International    Thermonuclear 
Experimental Reactor tokamak, estimated at $10^{10}$ \$.

\section{ Conclusions }

In this work we have considered the  possibility  of  using 
the hypervelocity  impact  of  laser-compressed  pellets  as  an 
ignition mechanism of nuclear fusion processes. The acceleration 
of an incident projectile to velocities of about  $3.5  \times  10^6$   m 
s$^{-1}$  , which would result in an ignition temperature  of  about  4 
keV of the compressed deuterium-tritium fuel, can be achieved by 
positively charging the projectile to the  limits  permitted  by 
the mechanical rigidity  and  by  the  evaporation  fields.  The 
accelerating fields can be produced  either  by  a  low-$\beta$ linac 
operating   at   superconducting   temperatures,   or   by    an 
electric-field wavefront generated with the  aid  of  sparkgap 
switches successively activated by the moving  charged  capsule. 
An accelerator length of 13 km was estimated on  the  assumption 
that deuterium-tritium densities of 570 g cm$^{-3}$   could be obtained 
by laser-driven compression. 

The purpose of this paper was of introducing the concept of 
hypervelocity ignition of fusion reactions  in  laser-compressed 
fuel  targets.  The  analysis  of  this   work   suggests   that 
existing laser facilities used for inertial  confinement  fusion 
research may have the  capability  to  achieve  ignition  and  a 
moderate  gain  when  used  in  conjunction  with  hypervelocity 
acceleration. At the same time, the  development  of  MJ  driver 
energy remains a necessity for the  compression  of  large-yield 
targets. Detailed further work is needed however to  assess  the 
physical efficiency of the impact  ignition  mechanism,  and  to 
determine the parameters of the accelerating structure  and  the 
associated costs.

\newpage
Figure captions\\

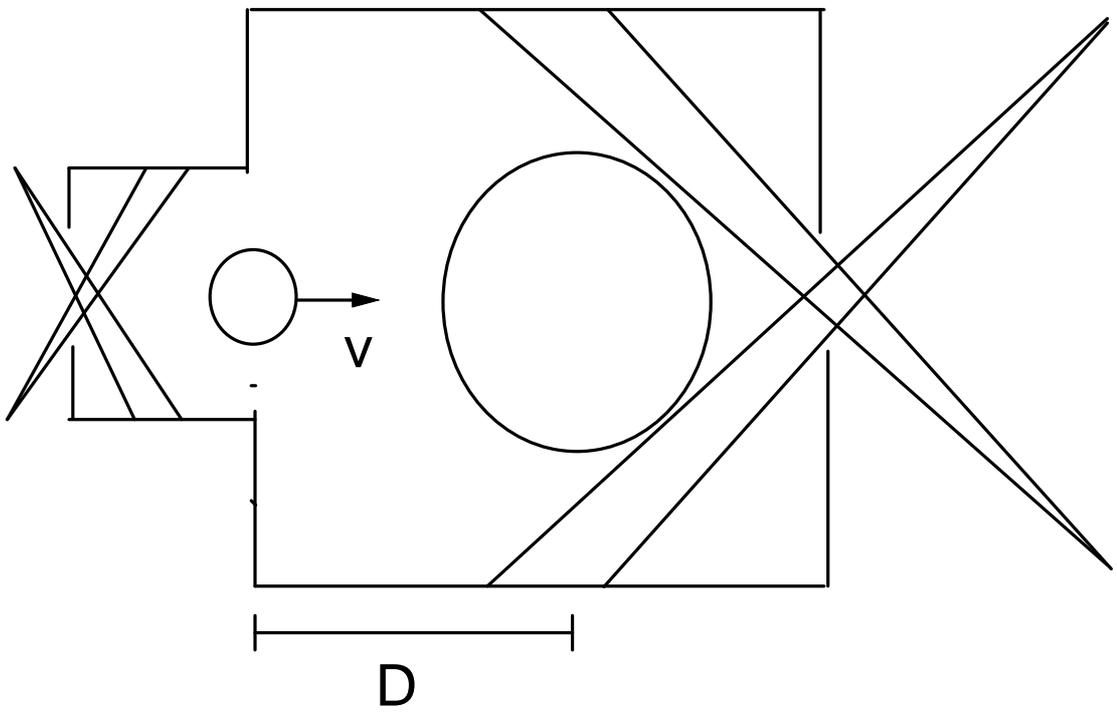
Fig. 1. Ignition of fusion reactions by the hypervelocity impact 
of an incident projectile of velocity $v$ with a large-yield  fuel 
target at rest. The collision  takes  place  inside  the  larger 
cavity of a double cylindrical high-$Z$  case,  which  converts  a 
laser pulse into X-rays which then drive the incident projectile 
and the large-yield target at rest into high-density states. The 
difference in the compression times of the incident capsule  and 
of the large-yield target is compensated by the  fact  that  the 
X-ray pressure is exerted on the incident  projectile  only  for 
the final part of its path.  A  second,  short  laser  pulse  is 
projected inside  the  smaller  cavity  to  ensure  the  uniform 
illumination of the incident projectile during its entrance into 
the  larger  cavity.  $D$  is  the  distance  from  the   incident 
projectile entrance into the large cavity to the center  of  the 
fuel target at rest.\\

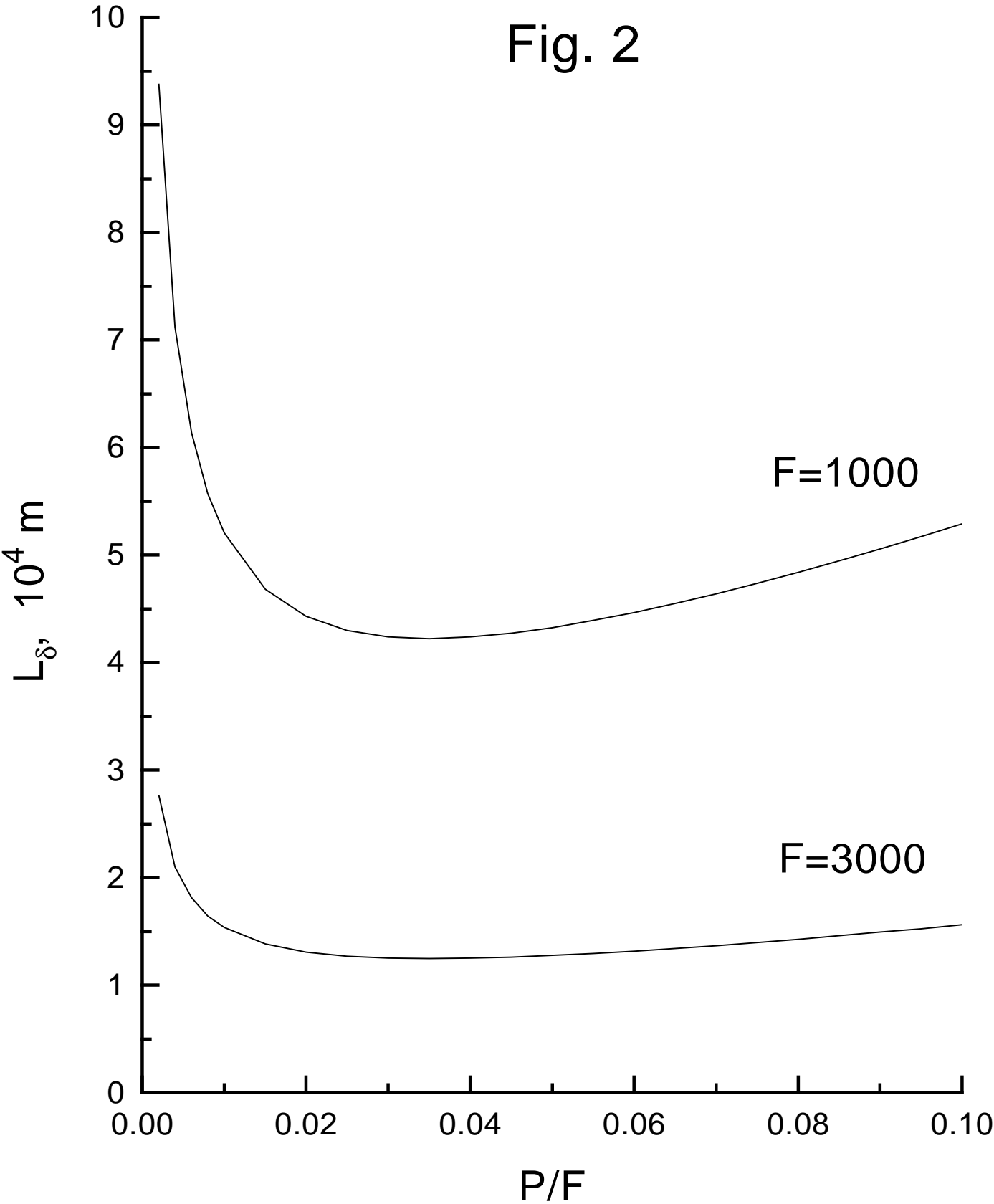
Fig. 2. Accelerator length $L_\delta$  as a function of  the  ratio  $P/F$, 
for the values  $F=1000$  and  $F=3000$.  It  is  assumed  that  the 
material of the projectile is beryllium, and $A=10, \eta=0.5, N=2$.\\

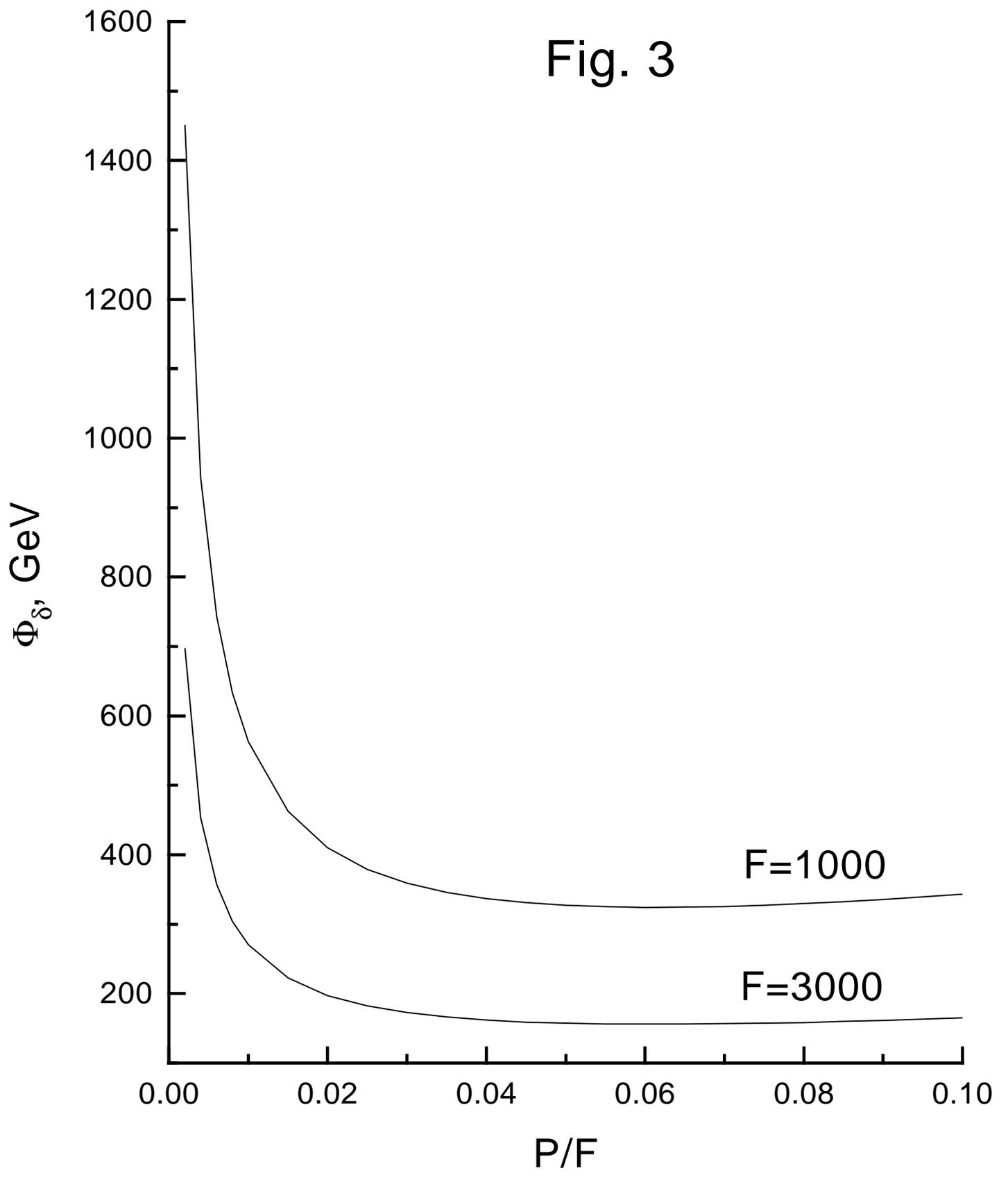
Fig. 3. Potential difference $\Phi_\delta$ as a function of the ratio  $P/F$, 
for the values  $F=1000$  and  $F=3000$.  It  is  assumed  that  the 
material of the projectile is beryllium, and $A=10, \eta=0.5, N=2$.\\

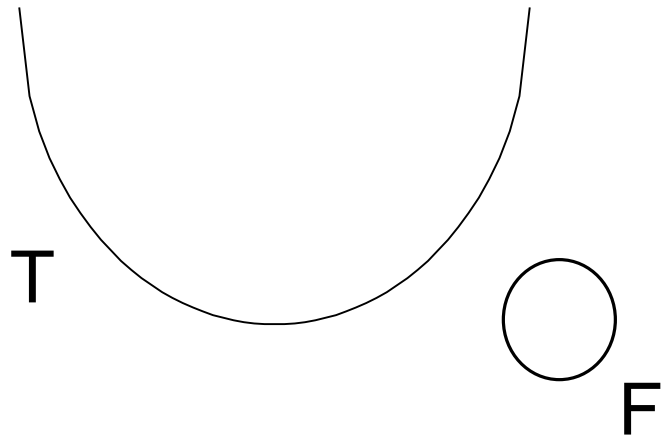
Fig. 4. Gas-insulated, three-electrode spark  gap  connecting  a 
drift tube to a positive high-voltage bar. The upper electrode T 
is connected to  the  drift  tube,  the  lower  electrode  A  is 
connected to the  high-voltage  bar,  and  F  is  lying  in  the 
proximity of T and has initially the same potential as  T.  When 
the positively-charged capsule is inside  the  drift  tube,  the 
electrode T becomes positively charged with respect  to  F,  and 
that initiates a spark across the close gap between T and F. The 
ionization  thus  created  then  initiates  the  main  discharge 
between T and A, at the  end  of  which  the  potential  of  the 
electrode T and of the drift  tube  to  which  it  is  connected 
becomes highly positive.

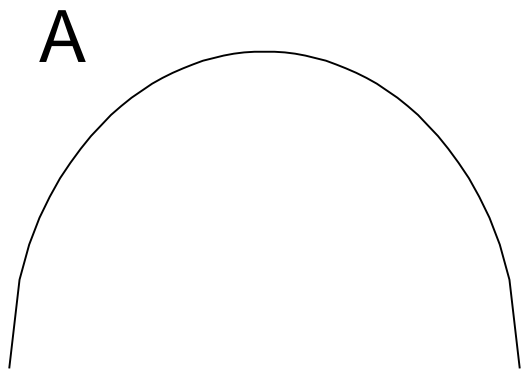
\end{document}